\documentclass{article}
\usepackage{siunitx}
\usepackage{arxiv}
\usepackage{xurl}
\usepackage{booktabs}
\usepackage[colorlinks=true,allcolors=blue]{hyperref}
\usepackage[all]{hypcap}
\usepackage[utf8]{inputenc} % allow utf-8 input
\usepackage[T1]{fontenc}    % use 8-bit T1 fonts
\usepackage{url}            % simple URL typesetting
\usepackage{booktabs}       % professional-quality tables
\usepackage{amsfonts}       % blackboard math symbols
\usepackage{nicefrac}       % compact symbols for 1/2, etc.
\usepackage{microtype}      % microtypography
\usepackage{doi}
\usepackage{makecell}
\usepackage{amsmath}
\usepackage[numbers]{natbib}
\usepackage[capitalise,noabbrev]{cleveref}
\usepackage{graphicx}
\usepackage{cuted}
\graphicspath{ {./images/} }
\usepackage{xcolor}
\usepackage{comment}
\usepackage[acronym]{glossaries}
\usepackage{fancyhdr} % Used to add line in header
\glsdisablehyper  % Use this to remove all the hyperlinks on every acronym
\usepackage{gensymb}

\newcommand{\affit}[1]{$^{\mathrm{\textnormal{\textit{#1}}}}$}
\newacronym{ecmwf}{ECMWF}{European Centre for Medium-Range Weather Forecasts}
\newacronym{ddms}{DDMs}{Data-driven models}
\newacronym{ddm}{DDM}{data-driven model}
\newacronym{eps}{EPS}{ensemble prediction system}
\newacronym{epss}{EPSs}{ensemble prediction systems}
\newacronym{metno}{MET Norway}{Norwegian Meteorological Institute}
\newacronym{metcoop}{MetCoOp}{Meteorological Co-operation on Operational \acrshort{nwp}}
\newacronym{nwp}{NWP}{numerical weather prediction}
\newacronym{synop}{SYNOP}{surface synoptic observations}
\newacronym{nmhses}{NMHSes}{national meteorological and hydrological services}
\newacronym{gnns}{GNNs}{graph neural networks}
\newacronym{gnn}{GNN}{graph neural network}
\newacronym{gpu}{GPU}{graphical processing unit}
\newacronym{gpus}{GPUs}{graphical processing units}
\newacronym{lam}{LAM}{limited area model}
\newacronym{lams}{LAMs}{limited area models}

\newacronym{t}{T}{2-meter temperature}
\newacronym{ws}{WS}{10-meter wind speed}
\newacronym{p6h}{P6h}{6-hour precipitation accumulation}
\newacronym{mslp}{MSLP}{mean sea level pressure}
\newacronym{tmin}{Tmin}{daily minimum 2-meter temperature}
\newacronym{tmax}{Tmax}{daily maximum 2-meter temperature}
\newacronym{wsmax}{WSmax}{daily maximum wind speed}
\newacronym{p24h}{P24h}{24-hour precipitation accumulation}

\newacronym{ifs}{IFS}{Integrated Forecast System}
\newacronym{meps}{MEPS}{MetCoOp Ensemble Prediction System}
\newacronym{aifs}{AIFS}{Artificial Intelligence Forecasting System}

\newacronym{fft}{FFT}{fast Fourier transform}
\newacronym{relu}{ReLU}{rectified linear unit}
\newacronym{mlp}{MLP}{multilayer perceptron}
\newacronym{nlp}{NLP}{Natural Language Processing}

\newacronym{crps}{CRPS}{Continuous Ranked Probability Score}
\newacronym{fcrps}{fCRPS}{Fair Continuous Ranked Probability Score}
\newacronym{ssr}{SSR}{spread-skill ratio}
\newacronym{mse}{MSE}{mean squared error}
\newacronym{mae}{MAE}{mean absolute error}
\newacronym{rmse}{RMSE}{root mean square error}
\newacronym{ets}{ETS}{equitable threat score}
\newacronym{fss}{FSS}{fractions skill score}
\newacronym{dct}{DCT}{discrete cosine transform}
\newacronym{bss}{BSS}{Brier skill score}
\newacronym{qq}{QQ}{quantile-quantile}

\title{Exploring the role of input data on hail nowcast skill using spatiotemporal neural networks}

\author{
 George Pacey \affit{a} \\
  \And
  Oph\'elia Miralles \affit{b} \\
  \And
Ulrich Hamann \affit{c} \\
\And
Olivia Martius \affit{a} \\
}

\begin{document}

\maketitle
\bigskip
\begin{abstract}
Hail can cause large financial losses and poses risks to public safety, making reliable nowcasts essential for timely warnings. Deep-learning approaches have emerged as a strong alternative to conventional methods, but how input data choices affect performance has not been deeply explored. We investigate how the skill of a deep-learning hail nowcasting model can be improved without changing the model architecture. Sensitivity experiments assess the impact of training-data volume, random data augmentation (mirroring and rotation) and the number of input timesteps. Increasing the years of training data substantially improves forecast skill, by up to 25 minutes at later lead times. Augmentation improved performance for larger datasets but interestingly degraded performance for smaller ones. Sensitivity to input timesteps was weaker than sensitivity to training years. These findings show that improvements in data selection and preprocessing can yield substantial gains even with a fixed architecture, offering guidance for future deep-learning nowcasting development.
\end{abstract}

\bigskip
\vspace{1.5cm}

\footnotetext[1]{Institute of Geography, Oeschger Centre for Climate Change Research, University of Bern, Bern, Switzerland}
\footnotetext[2]{Norwegian Meteorological Institute, Oslo, Norway}
\footnotetext[3]{MeteoSwiss, Locarno, Switzerland}
\footnotetext{*Corresponding author: \texttt{george.pacey@unibe.ch}}

\section{Introduction}
Convective hazards, including hail, lightning and extreme precipitation cause substantial economic losses and pose a direct threat to public safety \citep{kopp2023summer}. In Switzerland, individual hailstorms can generate losses of several hundred million Swiss francs \citep{GVL_2022}. Similar financial losses have also occurred in neighbouring countries \citep{Kunz_et_al_2018}. These impacts on infrastructure, agriculture, cars and buildings highlight the need for accurate and timely short-range forecasts (nowcasts) to support warnings for sectors such as aviation, emergency management and the general public. Despite their societal relevance, convective hazards remain difficult to predict because of their highly non-linear development and rapid evolution. Deep-learning models have emerged as a promising alternative to traditional nowcasting methods \citep{Shi_et_al_2015, Franch_et_al_2020, Espeholt_et_al_2022, Miralles_et_al_2025}, although precipitation has attracted more attention in previous literature compared to convective hazards. At MeteoSwiss, the COALITION-4 model was developed to predict the probability of lightning, hail, and precipitation over Switzerland using a combination of convolutional layers and recurrent units \citep{Leinonen_et_al_2023, Rombeek_et_al_2024}. The model runs at 1~km spatial resolution and produces nowcasts up to 1~hour ahead. The model used the year 2020 as training data and was shown to outperform the conventional motion-vector extrapolation approach \citep{Hering_et_al_2008} when using the model for thunderstorm warnings \citep{Hamann_et_al_2025}. COALITION-4 has been used operationally at MeteoSwiss since summer 2026 for thunderstorm warnings, including applications in the MeteoSwiss app, civil protection, and the Joint Information Platform for Natural Hazards. Operational data-driven nowcasting systems are also in development and being implemented at other European weather services \citep{boloni_et_al_2025,montmerle_et_al_2025} and also in the United States (see review in \citealt{McGovern_et_al_2023}). %Related work in the United States has also explored UNet-based thunderstorm nowcasting using satellite data \citep{Ortland_et_al_2023}. 

Despite encouraging advances, deep-learning nowcasting models still have important limitations. Skill typically decreases rapidly with lead time and predictions can become overly smooth. Hail is a particularly challenging hazard to model; \citet{Leinonen_et_al_2023} showed that the probability of hail model had a lower accuracy than the probability of lightning model, especially at later lead times. Given the growing volume of archive data, an open question is how increasing the size of the training dataset affects model skill. Enriching the training dataset can also be performed through artificial data augmentation strategies such as random rotation, mirroring and cropping of 2D tiles during training. These strategies were already applied during the original development of convolutional neural networks \citep{LeCun_et_al_1998} and have been applied to various meteorological nowcasting applications  \citep{Tran_and_Song_2019,Lagerquist_et_al_2020, Leinonen_et_al_2022,Leinonen_et_al_2023, Rombeek_et_al_2024, Franch_et_al_2025}. When using spatiotemporal neural networks for nowcasting applications, applying mirroring and rotations to the training data results in loss of information about the climatologically most frequent direction of thunderstorm motion. Instead, all directions of motion occur with roughly the same frequency. Whether the augmentation strategy is effective and optimal across different applications merits further exploration. Additional input data choices must also be made such as the tile size (e.g. 256 x 256 pixels) and the number of input timesteps. 
%Moreover, most previous deep-learning nowcasting studies have focused on precipitation, while comparatively fewer studies have addressed convective hazards such as hail and severe wind gusts. These hazards can cause substantial economic damage and pose a direct risk to life. In Switzerland, the severe thunderstorm wind event in La Chaux-de-Fonds recently illustrated this risk, causing damage estimated between 70 and 90 million francs and one fatality \citep{MeteoSwiss_Chaux-de-Fonds}.

Here, we explore the role of input data on skill of a deep-learning hail nowcasting model while keeping the model architecture fixed. The focus is on the number of training years, data augmentation choices and the number of input timesteps. This is achieved by performing a series of sensitivity experiments. Section~\ref{sec:data} describes the data used for training the model. Section~\ref{sec:model_setup} introduces the model setup. Section~\ref{subsec:sensitivity_tests} shows the sensitivity tests and a deeper intermodel comparison is performed in Section~\ref{sec:final_model_val}. Section~\ref{sec:conclusion} summarises the findings of the current study. 

\section{Data}\label{sec:data}
%The present study does not include NWP inputs, in order to first assess the impact of training-data volume and preprocessing choices using radar and static topographic information only. 

\subsection{Input variables}

The model in this study takes as input radar-based variables on the Swiss coordinate system grid (LV95) \citep{swisstopo_2024} at a 1 km spatial resolution. Compared to the model in \citet{Leinonen_et_al_2023}, the predictor set is simplified to three radar variables based on an ablation study, which builds on \citep{Leinonen_et_al_2023}'s exploration of the importance of different data sources (radar, satellite, lightning). Since the model did not rely heavily on several individual predictors (Figure S1), only three radar variables are used in this study to lower model complexity: probability of hail (POH), maximum column radar reflectivity (CZC) and echo-top height of 20~dBZ (EZC20). In Section~\ref{subsec:dataset_size}, these are provided from the previous 8 timesteps (40~minutes) whereas the number of input timesteps is varied in Section~\ref{subsec:num_input_timesteps}.

%Swiss meteorological datasets are increasingly available through Open Government Data (OGD; \citep{Meteo_Swiss_OGD}). These includeradar-derived variables relevant for model training, such as the probability of hail (POH) and the maximum expected severe hail size (MESHS). POH and MESHS are produced from the volumetric radar reflectivity following \citet{nisi2016spatial} and provide proxies for the probability of hail occurrence and hail size at kilometer-scale resolution. In this study, we use the MeteoSwiss POH product for the convective season May--September.

\subsection{Target variable}

The target is POH (probability of hail) for the 12 future timesteps, corresponding to 60 minutes. POH is a radar-based hail proxy that uses the distance between the maximum height of the 45 dBZ radar echo and the freezing level to infer the probability of hail at the surface. The calibration in Switzerland was performed against hail pad observations by \citet{Foote_et_al_2005} based on data from \citet{Waldvogel_et_al_1979}. For further details, we refer to \citep{Kopp_et_al_2024}. 

%It was showed that according to quality-controlled crowdsourced reports that hail can occur when POH=0\% \citep{Kopp_et_al_2024}, though most reports are of small hailstones around 1 cm or below, while thunderstorm warnings at MeteoSwiss are designed to be triggered when hail diameter exceeds 2 cm. Despite its limitations, using POH as the model ground truth enables a direct comparison with the baseline model. Furthermore, POH is available in the full Swiss radar domain with archived data since 2011. For further details, we refer to \citep{Kopp_et_al_2024}. 

\subsection{Tile size selection}
\label{sec:tile_size}

Operationally, the COALITION-4 model runs on a 352 × 448 grid at 1 km resolution. Training with this tile size is RAM-intensive and exacerbates class imbalance because hail cells have a small spatial extent. A larger tile contains a much higher proportion of zero-valued grid points. Although the model should predict hail non-occurrence correctly, it should not overemphasise this task. Powers of two are commonly used for tile sizes, such as $256\times256$, because they permit up to eight downsampling steps but this depth is rarely necessary. We opt for $160\times160$ tiles to reduce class imbalance and to allow the model to focus on hail-cell evolution. It also provides a sufficient spatial extent since assuming a cell originates at the tile centre and moves eastward, it would need to travel at 128 km hr\textsuperscript{-1} on average to exit a $256\times256$ tile, a speed that is extremely rare \citep{Wapler_and_James_2015}. A $160\times160$ tile still permits up to five downsampling steps if required.

\subsection{Construction of the training samples}\label{subsec:construct_samples}

Training tiles are selected by first identifying timesteps with a contiguous region where POH exceeds 25\% over at least 15~km$^2$, initially centred on the largest POH region. After extending 80 pixels (kilometres) north, east, south and west, tiles extending more than 20 pixels beyond the domain are discarded; otherwise their coordinates are clipped to the domain limits. Therefore, the hail feature is not always exactly centred. To limit overlap on the same timestep, candidate tiles are kept only if centres are at least half a tile width (80~km) apart. Temporal sequences are then extracted for each tile, the input is the eight timesteps preceding the current frame (40~minutes at 5-minute resolution) and the target starts at the current timestep and extends 12 timesteps ahead. $X$ and $y$ are thus spatiotemporal blocks of shape time~$\times$~height~$\times$~width~$\times$~channel(s), with $X$ having 3 input channels (POH, CZC and EZC20) and $y$ one output channel (future POH). We call each spatiotemporal sequence a \textit{sample}. Training batch size is 48 samples.

%Following \citet{Leinonen_et_al_2023}, we retain the probabilistic nature of POH and train the model using a pointwise binary cross-entropy (BCE) loss. We also tested a Brier score loss, but it resulted in slightly poorer calibration than BCE, though not significant. Using BCE allows a more direct comparison with the baseline model.

\subsection{Dataset splitting}

The years 2019--2020 and 2022--2025 are used for training and validation, split so that timesteps from the same or adjacent days are never assigned to different subsets. This reduces temporal dependence between the two datasets, making validation metrics monitored during training a more reliable indicator of generalisation. A similar climatological hail probability in both subsets is also ensured by randomly shuffling days between the two sets until the climatologically hail probability was within 1\% for both. The year 2021 is held out entirely for testing to robustly assess performance on unseen data uncorrelated with the training or validation set. The year 2021 was a particularly notable year for hail in Switzerland with record damage totals \citep{GVL_2022} and a large number of supercell occurrences, which are especially important to forecast given their association with increased hazard risk \citep{Markowski_Richardson_2010}. The splitting results in 81\%, 6\% and 13\% of samples being assigned to the training, validation and testing sets, respectively. The validation set is typically used to tune hyperparameters and adjust model architecture; here it is used only to monitor training and select the best model checkpoint. The sensitivity test models (Section~\ref{subsec:sensitivity_tests}) are applied to both the validation set (climatologically consistent but used for checkpoint selection) and the test set (an unusual convective season, fully independent).

\section{Model and experimental setup}
\label{sec:model_setup}
\subsection{Model architecture}

The model architecture follows \citet{Leinonen_et_al_2022, Leinonen_et_al_2023} combining a U-shaped network (U-Net; \citealp{Ronneberger_et_al_2015}) with a gated recurrent unit (GRU; \citealp{Ballas_et_al_2016}). Combining the UNet and GRU (hereafter UNetGRU) enables joint learning of spatial and temporal patterns. Similar architectures appear in other nowcasting studies \citep{Franch_et_al_2020, Espeholt_et_al_2022, Andrychowicz_et_al_2023}. The only difference from \citet{Leinonen_et_al_2023} is the number of convolutional filters per UNet resolution level -- 24, 48, 96 here versus 32, 64, 128 -- reflecting this study's reduced input variables. This lowers learnable parameters from $\sim$6.5 to 2.5 million and reduces compute time. For further details on the architecture see Section~3a of \citet{Leinonen_et_al_2022}.

\subsection{Random data augmentation}\label{subsec:random_aug}

During training, one of eight transformations is randomly applied to each sequence as it enters a batch: no change (original), reflection about the horizontal, vertical or two diagonal axes, or rotation by 90, 180 or 270 degrees, each with 12.5\% probability of being chosen. The effect of each transformation on a north-easterly propagating cell is shown in Figure S2. The validation and testing datasets are left unchanged. Hereafter, random data augmentation is referred to as "augmentation" or simply "aug". 

\subsection{Loss function, learning rate schedule and early stopping}

Following \citet{Leinonen_et_al_2023}, binary cross entropy (BCE) is used as the loss function against the probabilistic target POH. The learning rate starts at 0.001 and is multiplied by 0.2 if validation loss does not improve for 5 epochs. Training is stopped (early stopping) after 8 epochs without improvement.

\section{Sensitivity experiments}\label{subsec:sensitivity_tests}

The sensitivity to increasing real data in the form of adding more training years is first explored. Second, the sensitivity to artificially increasing the diversity of training samples through random mirroring and rotation is analysed (see Section \ref{subsec:random_aug}). Finally, the sensitivity to the number of input timesteps is analysed to explore how increasing the temporal context affects model performance.

\subsection{Training dataset size}\label{subsec:dataset_size}

\begin{figure}[h]
\centering
\includegraphics[width=.95\textwidth]{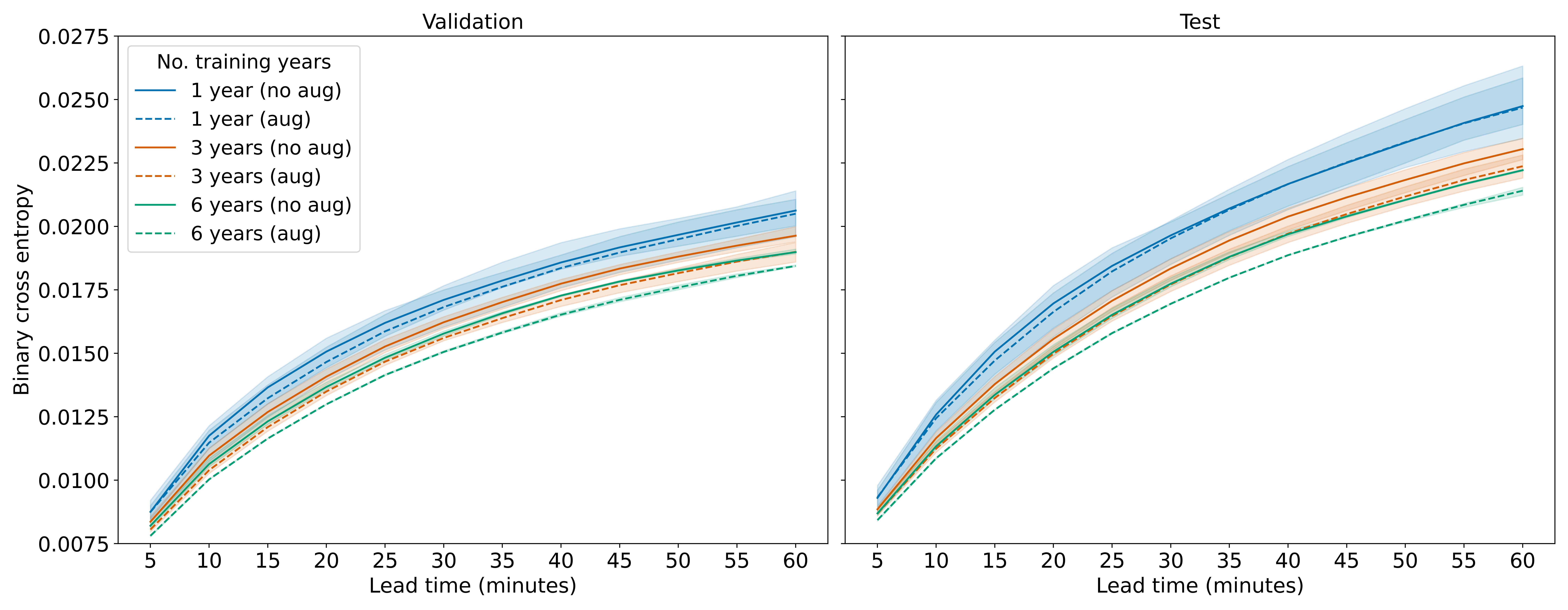}
\caption{BCE vs lead time when using 1, 3 or 6 years of training data. The solid and dashed lines show experiments without or with augmentation, respectively. The shading represents the range of results depending on the combination of years and running the same experiment 3 times with a different batch order. The full list of sensitivity experiments is shown in Table S1.}
\label{fig:BCE_leadtime_aug_year}
\end{figure}

Model performance improves with increasing training data both with and without augmentation (Figure \ref{fig:BCE_leadtime_aug_year}). The improvement is larger from 1 to 3 years than from 3 to 6 years, suggesting a possible plateau. The benefit of additional training data is more pronounced at longer lead times. For example, the 6-year (aug) experiment achieves the same skill at 60 minutes as the average of the 1-year experiments at approximately 35 minutes. The best performance is achieved using 6 years of data with augmentation, while 3 years with augmentation is comparable or superior to 6 years without augmentation. These results are consistent across both the validation and testing datasets, although BCE is higher for the latter, primarily due to its higher climatological POH frequency.

\begin{figure}[h]
\centering
\includegraphics[width=.95\textwidth]{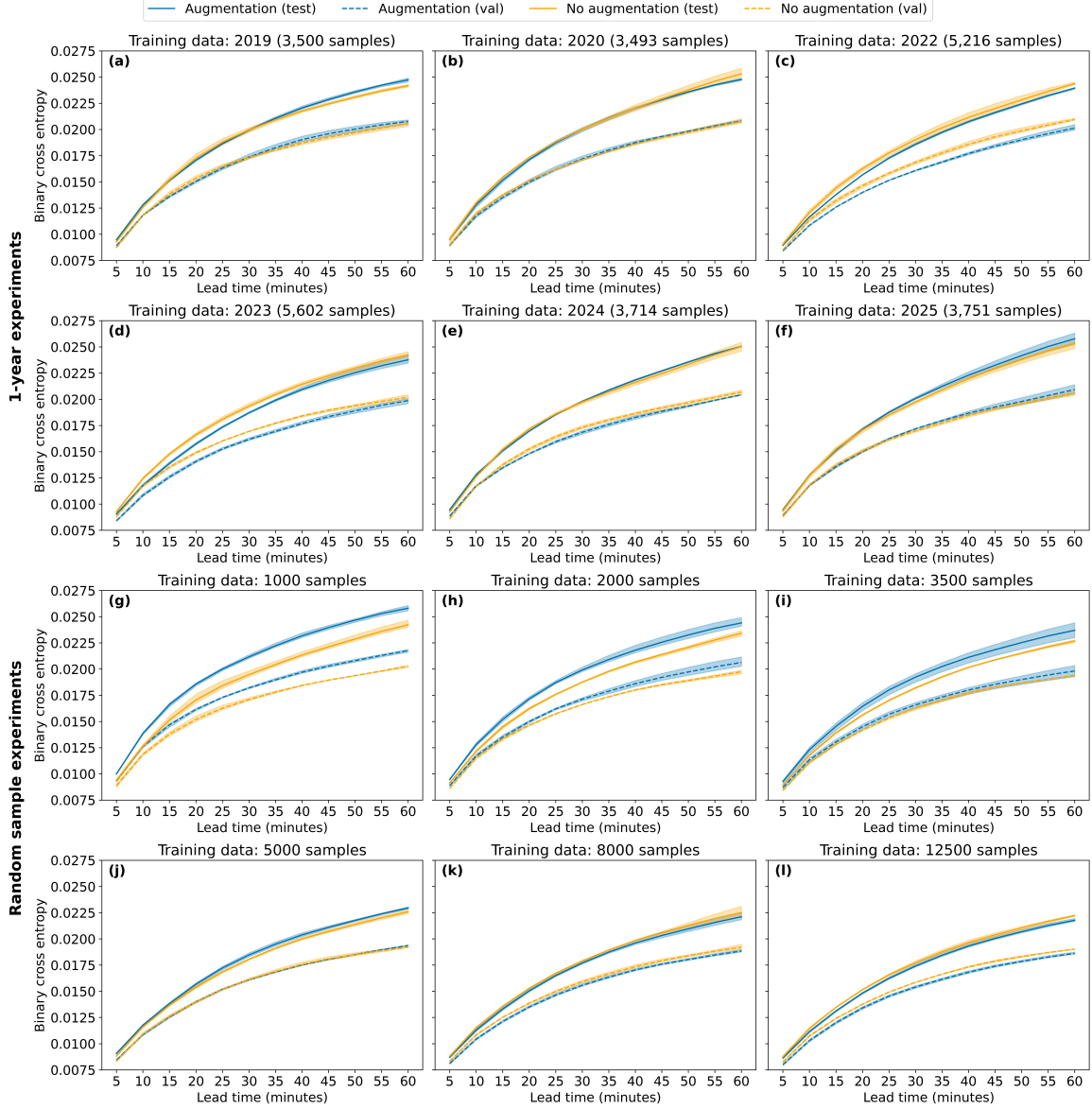}
\caption{As Figure \ref{fig:BCE_leadtime_aug_year}, but showing the individual one year experiments (a-f) and using different numbers of random selected samples from the six training years (g-l). The blue and orange lines show BCE with and without augmentation, respecitively. Solid lines represent the testing set while the dashed lines represent the validation set.}
\label{fig:BCE_one_year_exps_and_num_samples}
\end{figure}

The 1-year experiments show substantial overlap between the augmentation and non-augmentation experiment distributions (blue lines on Figure \ref{fig:BCE_leadtime_aug_year}) reflecting differences between the year used and batch ordering across three runs. Augmentation slightly degrades performance on average for some years, notably 2019 and 2025, but provides the largest improvement in 2022 and 2023, which have the largest sample counts (Figure \ref{fig:BCE_one_year_exps_and_num_samples}a-f).

To further understand the effectiveness of augmentation depending on the number of training samples, 1,000, 2,000, 3,500, 5,000, 8,000, 12,500 samples were randomly sampled from the total 25,276 available in the 6 year training dataset (Figure \ref{fig:BCE_one_year_exps_and_num_samples}g-l). According to these sensitivity experiments, augmentation slightly improves performance for larger sample counts but interestingly degrades performance for increasingly small training sets. It should be noted that the effect of augmentation also depends on the nature of samples within a training dataset. In particular for 1-year training datasets, the effects vary considerably. The years 2022 and 2023 had the largest number of samples and augmentation was most effective, however the effect of augmentation was less significant for other years that had a comparable sample count. There is substantial interannual variability in hail occurrence, including differences between northern and southern Switzerland \citep{Wilhelm_et_al_2024}.

The result that augmentation degrades performance with decreasing sample size is counterintuitive, since neural networks usually generalise better with larger and more diverse training data \citep{LeCun_et_al_1998}. We provide a hypothesis for this result. For our application, the mirroring and rotation strategy balances the storm propagation directions seen during training. In Switzerland, convective systems commonly propagate eastward (north-easterly, easterly or south-easterly), while westerly propagation is less common due to the prevalence of westerly flow \citep{nisi2018alpine}. Balancing the propagation directions seen during training may avoid the model from overfitting to the easterly propagation direction. However, with a low sample count the model may less effectively learn easterly propagation as it sees limited samples of such cases. It is important that the model learns this propagation direction effectively as it is the most common propagation direction. With more diversity of real cases (e.g. 6 years), despite the balancing of cell motion directions, the model still sees sufficient easterly propagating systems. 
%These results highlight that commonly applied augmentation with flipping and rotation may not always improve generalisation. Future work may benefit from exploring different data augmentation strategies that increase training diversity while not significantly changing the climatological distribution of motion directions. 

Using 3,500 and 12,500 samples drawn from 6 years allows us to compare the benefit of overall sample count versus sample diversity, since more years means storms of different intensities, propagation speeds and directions. Each of 2019, 2020, 2024 and 2025 has approximately 3,500 samples. Randomly drawing this many samples from all six years performs significantly better than these 1-year experiments. Likewise, randomly sampling 12,500 samples (approximate count in 3-year experiments) from the 6-year dataset highlight the same positive effect of increased diversity. The best 1,000 sample experiments also sometimes outperform the 1-year experiments. These results highlight the value of increasing real training data diversity even at constant sample size, consistent with a study on post-processing high-resolution NWP data to infer lightning probability \citep{Yousefnia_et_al_2024}.

\subsection{Number of input timesteps}\label{subsec:num_input_timesteps}

\begin{comment}
\begin{figure}[h]
\centering
\includegraphics[width=0.6\textwidth]{figures/BCE_leadtime_sensitivity_num_timesteps.png}
\caption{As Figure \ref{fig:BCE_leadtime_aug_year} but varying the number of input timesteps between 2 and 8. Six years of training data with augmentation is fixed for all experiments. The experiment with 3 years of data (8 input timesteps) and random augmentation is shown for reference in purple.}
\label{fig:BCE_leadtime_num_timesteps}
\end{figure}
\end{comment}

Finally, sensitivity to the number of input timesteps is explored, fixing 6 years of training data with augmentation (the best configuration from Section~\ref{subsec:dataset_size}) to limit the number of combinations. Averaging across the three runs, gave BCE loss of 0.01482, 0.01466, 0.01449 and 0.01460 for 2, 4, 6 and 8 inputs timesteps, respectively. Performance improved from 2 to 6 input timesteps but slightly declined at 8, albeit with high run-to-run variability with 2 and 4 input timesteps (Table S1). All six-input-timestep runs outperformed the eight-timestep runs, suggesting an optimal temporal window beyond which older observations may add limited predictive value and extra model complexity.

Skill is more sensitive to training years than to input timesteps: 2 timesteps with 6 training years outperformed 8 timesteps with 3 training years across all runs (Table S1). This suggests the model generalises better with more training data than from a longer temporal input context. Global data-driven weather prediction models can predict atmospheric states up to 10 days ahead using just two input states \citep{Lam_et_al_2023, lang_et_al_2024_aif}. Achieving comparable skill with fewer input timesteps would enable operational nowcasting systems to be more robust to missing radar timesteps and also run faster. 

\section{Final model evaluation}\label{sec:final_model_val}
\label{sec:results}

\begin{comment}
\begin{table*}[ht!]
\centering 
\footnotesize
\caption{Model configurations for the three models used in the final model evaluation.}
\label{tab:model_config}

\begin{tabular}{@{}lccc@{}}
\toprule
Configuration & Best model & 1,000 samples without augmentation & 1,000 samples + augmentation \\
\midrule
Training years & 6 years (all 25,276 samples) & 6 years (1,000 random samples) & 6 years (1,000 random samples) \\
Data augmentation & Rotations and mirroring & None & Rotations and mirroring \\
Previous timesteps input & 6 & 8 & 8 \\
\bottomrule
\end{tabular}
\end{table*}
\end{comment}

Based on the sensitivity tests (Section \ref{subsec:sensitivity_tests}), the model using 6 years of training data with augmentation and 6 input timesteps is selected as the best model. To better illustrate the effect of applying augmentation with low sample count, this model is compared to the models from the sensitivity tests using a randomly selected 1,000 samples from the total 25,276 samples (Figure \ref{fig:BCE_one_year_exps_and_num_samples}g). The models are applied to the full COALITION-4 domain (\(352 \times448\) km). Since the UNet part of the network is fully convolutional, the models trained on \(160 \times160\) km tiles can be directly applied to a larger domain. 

\begin{figure}[h]
    \centering
\includegraphics[width=\textwidth]{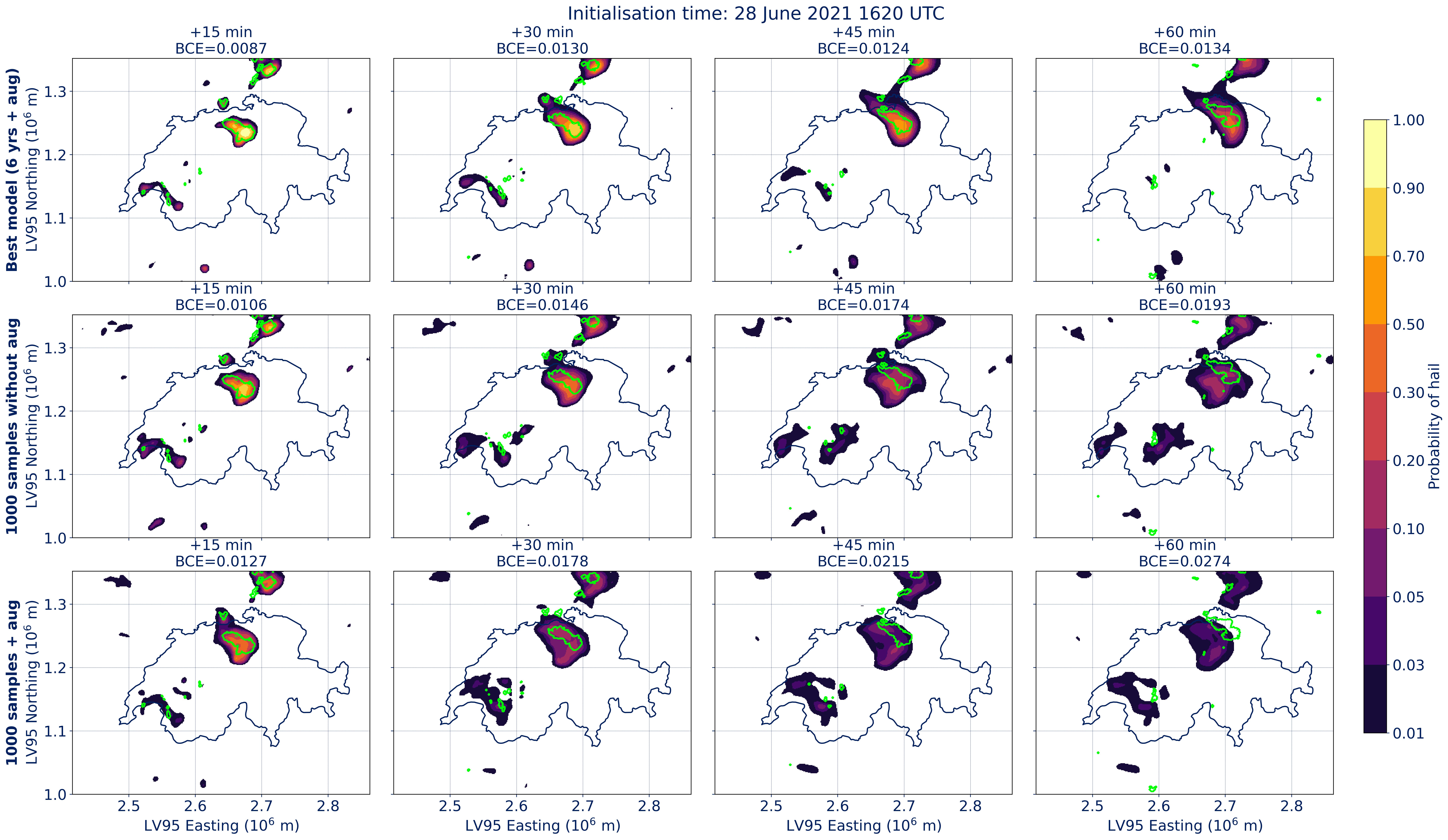}
    \caption{Nowcast initialised on 28 June 2021 at 1620 UTC. The top, middle and bottom rows show the 6 years with augmentation, 1,000 samples \textit{without} augmentation and 1,000 samples \textit{with} augmentation, respectively. The green contour shows the ground truth based on the 1\% probability contour. The colour scale indicates the predicted probability of hail, ranging from 1\% to 100\% POH. The value above each panel gives the BCE for that lead time.}
    \label{fig:case_study1}
\end{figure}

To understand general model behaviour, an example initialisation time is shown from a particularly impactful hail case in June 2021 \citep{kopp2023summer}. Figure~\ref{fig:case_study1} shows a POH nowcast sequence at lead times of 15, 30, 45 and 60 minutes. The main hail signal, initially just north of the domain centre, propagates north-eastward across the Swiss Plateau and crosses Zurich while another cell appears in the far north (southern Germany). At 15 minutes (first column), all models agree well with the ground truth in location, though the thousand-sample-with-aug model already produces lower probabilities over a slightly larger area, indicating lower confidence. This pattern continues to 30 minutes. By 45 minutes, this model's hail object is no longer centred on the ground truth and its probability area stops propagating. The thousand-sample-without-aug model shows this behaviour to a lesser extent, while the best model remains well aligned, producing areas with 50--70\% probabilities. As lead time increases, all models tend to lower hail probabilities while expanding the area exceeding 1\%, reflecting growing uncertainty. Both qualitatively and quantitatively (BCE in Figure~\ref{fig:case_study1} subfigure titles), this example reproduces the same model rankings as the sensitivity tests (Section~\ref{subsec:sensitivity_tests}). Augmentation degrades the 1,000-sample nowcast, likely because motion balancing reduced north-easterly, long-lived cells in training.

%The middle row in Figure~\ref{fig:case_study1} provides an intermediate comparison that helps separate the effect of adding more training years from other changes in the retraining setup. This model is trained only on 2020, as in the baseline, and uses the same random augmentation strategy but uses the retraining strategy (e.g. 160 x 160 tiles). This model produces probability fields that are closer to the final retrained model than to the baseline and reduces the broad high-probability areas produced by the baseline giving a more localised evolution of the hail signal. However, like the original baseline model there is a slight mismatch in the cell motion with increasing lead time. The final six-year model further improves the temporal coherence and spatial agreement compared to the ground-truth contour.

%The retraining approach using only 2020 has lower BCE than the baseline up to 45 minutes but becomes comparable to, or slightly worse than, the baseline at 50--60 minutes. Since this configuration differs from the baseline in several aspects of the model and training setup, the comparison indicates that changes other than the number of training years affect the score, but it does not isolate the contribution of any individual change.

\begin{figure}[h]
    \centering
\includegraphics[width=\textwidth]{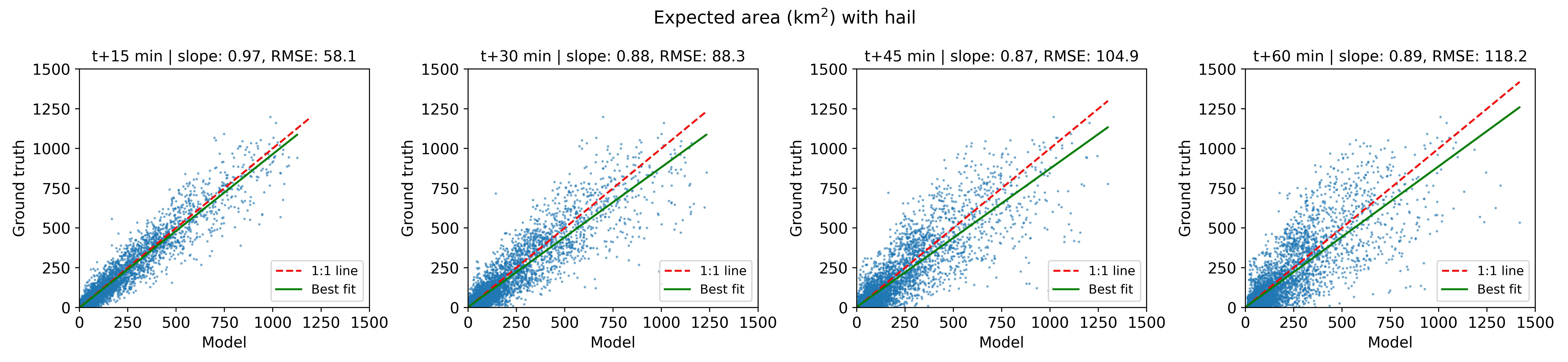}
    \includegraphics[
        width=\textwidth,
        trim=0 0 0 1.2cm, % left bottom right top
        clip
    ]{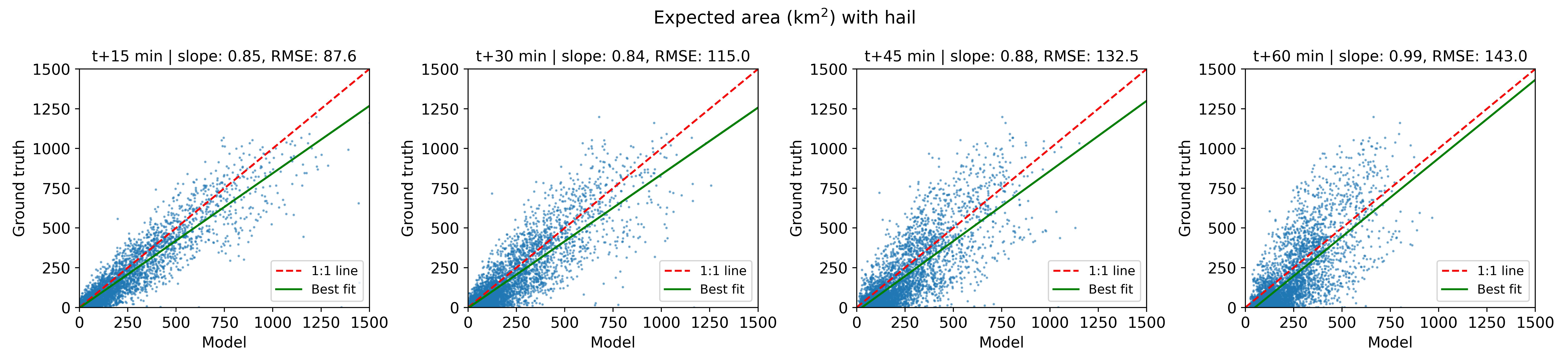}
    \includegraphics[
        width=\textwidth,
        trim=0 0 0 1.2cm, % left bottom right top
        clip
    ]{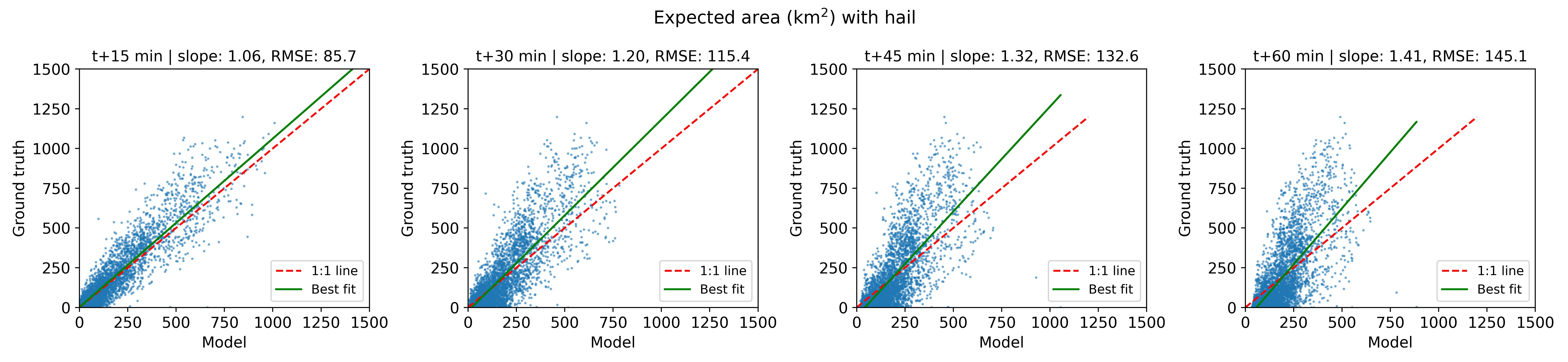}
    \caption{Expected hail area for nowcasts initialised at 5,363 times in 2021. Columns shows lead times of 15, 30, 45 and 60 minutes (left to right) and rows show the best model, 1000-sample without augmentation, and 1000-sample with augmentation (top to bottom). Each point is one nowcast initialisation time with model and ground-truth probabilities summed over the COALITION-4 domain (Figure \ref{fig:case_study1}) to give expected hail area in km$^2$. The dashed red line shows the 1:1 relationship and the green line the fitted linear relationship. Points above the 1:1 line indicate underestimation and points below indicate overestimation of hail-affected area.}
\label{fig:model_calibration_lead_time}
\end{figure}

Going beyond pointwise metrics, Figure~\ref{fig:model_calibration_lead_time} evaluates expected hail area across lead times. Since the ground truth is a radar-based proxy it will often differ from the actual hail area. Evaluation is carried out on nowcasts initialised from timesteps satisfying the tile-selection criterion (see Section \ref{subsec:construct_samples}), plus those within $\pm$30 minutes to include more instances before and after storm initiation.

The best model is well calibrated at 15 minutes lead time (top left), while the thousand-sample-without-aug model slightly overestimates (middle left) and the thousand-sample-with-aug model slightly underestimates (bottom left). The latter increasingly underestimates hail area with lead time, reaching a slope of 1.41 at 60 minutes. Nevertheless, case-by-case variability remains large and even the thousand-sample-with-aug model overestimates in some cases.

Reliability diagrams by lead time (Figure S3) show the best model close to the 1:1 line at 10 minutes, indicating near-perfect calibration. However, it becomes overconfident at later lead times, particularly above modelled probabilities of 0.2. For all models, jagged lines at higher probabilities reflect low sample counts in those bins, since the model rarely assigns such probabilities at these lead times. This jaggedness appears earlier for the thousand-sample-with-aug model, indicating a lack of confidence in higher probabilities already by 20 minutes. 

%The BCE as function of lead time was also calculated for the same 5,363 initialisation times over the full domain (Figure~\ref{fig:BCE_leadtime_full_domain}) showing the best model had the lowest BCE followed by the thousand-sample-without-aug model and thousand-sample-without-aug model (consistent with sensitivity tests).

\section{Conclusion}
\label{sec:conclusion}
This study investigates how the performance of a deep-learning hail nowcasting model can be improved without changing the underlying model architecture. In the framework of MeteoSwiss' deep learning nowcasting system (COALITION-4), we assessed the sensitivity of probabilistic hail nowcasts to the amount of training data, the use of random augmentation with rotation and mirroring and the number of input timesteps. The results show that data-related choices have a substantial impact on nowcast skill. 

The strongest and most robust improvement was obtained by increasing the number of training years. Training with six convective seasons substantially reduced the binary cross-entropy compared with training on a single year. With six convective seasons of training data, the model achieved comparable skill at 60 minutes lead time as the one-year experiments achieved at 35 minutes lead time. Additional experiments in which varying number of samples were drawn from six years of data showed that the diversity of the training data is important in itself, not only the total sample count. This indicates that exposure to a broader range of convective situations improves the model's ability to generalise to unseen cases.

The impact of random data augmentation through rotations and mirroring was more nuanced. While augmentation improved performance when enough real training samples were available, it generally degraded performance for smaller training sets. We hypothesise that this is related to the climatological distribution of storm motion. Mid-latitude convective systems typically propagate with an easterly component, whereas westerly propagation is less common. Random augmentation with rotations and mirroring artificially balances propagation directions and can therefore potentially help the model learn a wider range of possible motions. However, when the number of training samples is limited, this may reduce the effective number of examples representing the most common propagation directions, thereby degrading performance. These results suggest that random augmentation with transformations should not be assumed to improve generalisation for spatiotemporal nowcasting tasks and domain-knowledge guided augmentation strategies may be more optimal to mitigate overfitting.

The sensitivity to the number of input timesteps was weaker than the sensitivity to the number of training years. The best validation performance was obtained with six previous timesteps, corresponding to 30 minutes. However, models using fewer input timesteps still performed competitively when trained on the full six-year dataset. This suggests that for this task, additional training diversity is more valuable than a longer input sequence. This also has practical implications for operations, since models requiring fewer input timesteps may be more robust during periods with missing radar frames and may reduce inference cost.
\clearpage

\section*{Open data}

\noindent MeteoSwiss datasets are increasingly being made available through Open Government Data (OGD) \citep{Meteo_Swiss_OGD}. This includes the POH radar-based hail proxy used in this study. The model architecture code was already made available by previous COALITION-4 studies \citep{Leinonen_et_al_2023}: \url{https://github.com/MeteoSwiss/c4dl-multi}. The results, data and scripts of the sensitivity analysis and final evaluation in this study are available at the following links \url{https://doi.org/10.5281/zenodo.22131494} and \url{https://github.com/gpacey/C4-UNetGRU-sens-exps}

\section*{Acknowledgements}

We acknowledge a preparatory project and small project grant under the Swiss AI initiative for providing compute resources at the Swiss National Supercomputing Centre (CSCS). We thank Matteo Buzzi for insights into MeteoSwiss' datasets and the COALITION-4 model as well as Martin Aregger for discussions about MeteoSwiss radar datasets and for providing a radar processing script. G.Pacey and O.Martius acknowledge funding from the Mobiliar. The funding provider did not play any role in the study. 

\section*{Computational Resources}

Training utilised resources from the Swiss National Supercomputer Centre (CSCS). All experiments were run in TensorFlow inside an NVIDIA aarch64-compatible container and distributed over four NVIDIA H200 GPUs via batch data parallelism. Wall-clock training times varied across sensitivity experiments (Table \ref{tab:sensitivity_exps}), with the most computationally expensive configuration (six years of training data, eight input timesteps and augmentation) completing in approximately 4 GPU hours (1 node hour with 4 GPUs). 

\clearpage

\appendix

\renewcommand{\thetable}{A\arabic{table}}
\renewcommand{\theHtable}{A\arabic{table}}
\setcounter{table}{0}

\renewcommand{\thefigure}{A\arabic{figure}}
\renewcommand{\theHfigure}{A\arabic{figure}}
\setcounter{figure}{0}

\begin{figure}[h]
    \centering
\includegraphics[width=0.65\textwidth]{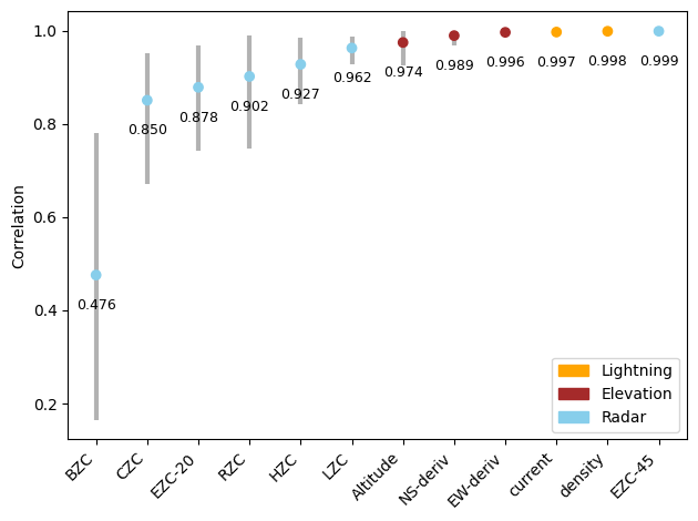}
    \label{fig:baseline_model_ablation}
    \caption{Input feature importance of the POH model developed by \citep{Leinonen_et_al_2023} based on permutation feature importance. High correlation indicates little difference between the permuted and non-permuted nowcast indicating low importance of that feature. Low correlation indicates high importance of that feature. 
}
\end{figure}

\begin{figure}[h]
    \centering
\includegraphics[width=0.95\textwidth]{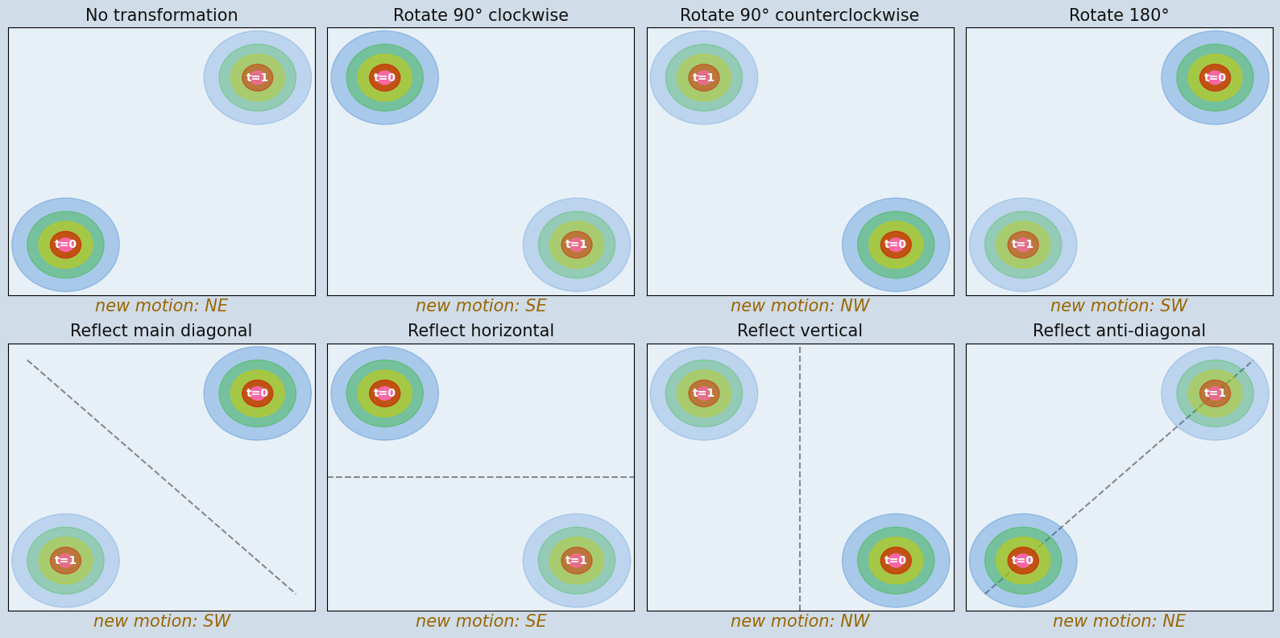}
    \caption{Schematic of the 8 possible transformations with the augmentation strategy shown for an example with a cell moving from south-west to north-east. Note that due to the positioning of the example cell certain transformations have the exactly same effect (e.g. rotated 90 degrees and reflect vertical). However this would not be the case when cells are not evenly displaced from the vertical and horizontal axes. 
}
    \label{fig:random_aug_examples}
\end{figure}

\begin{figure}[h]
    \centering
\includegraphics[width=\textwidth]{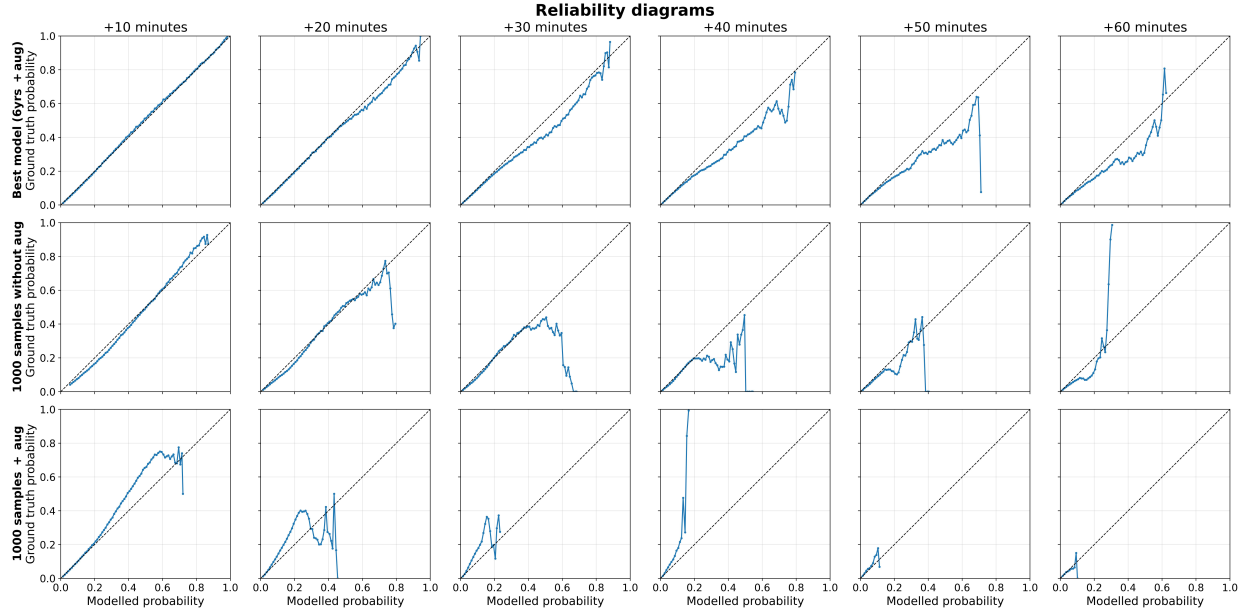}
    \caption{Reliability diagrams for the best model, thousand-sample-without-aug and thousand-sample-with-aug models as a function of lead time. The dashed diagonal denotes perfect reliability.}
    \label{fig:reliability_diagram}
\end{figure}

\begin{table}[t]
\centering
\footnotesize
\caption{Summary of sensitivity experiments. The loss is averaged across all lead times; lower BCE is better. Three runs are shown that use the same initial weights but different batch order. For the augmentation experiments, the type of augmentation is random, meaning the total number of a given augmentation type applied could vary slightly between experiments. Bold indicates the better mean loss between without and with augmentation; a star ($^*$) indicates all three runs are better. All experiments use 8 input timesteps unless otherwise stated. A sample refers to a spatiotemporal sequence. The value highlighted in green shows the best model across all sensitivity experiments.}
\label{tab:sensitivity_exps}
\begin{tabular}{@{}cccll@{}}
\toprule
Training year(s) & Samples & Input Timesteps & Validation loss without aug & Validation loss with aug \\
\midrule
2019                   & 3,500  & 8 & \textbf{0.01675, 0.01660, 0.01665} & 0.01695, 0.01660, 0.01677 \\
2020                   & 3,493  & 8 & 0.01677, 0.01656, 0.01659          & \textbf{0.01662, 0.01653, 0.01677} \\
2022                   & 5,216  & 8 & 0.01646, 0.01658, 0.01631          & \textbf{0.01568, 0.01585, 0.01570}$^*$ \\
2023                   & 5,602  & 8 & 0.01635, 0.01632, 0.01645          & \textbf{0.01587, 0.01559, 0.01567}$^*$ \\
2024                   & 3,714  & 8 & 0.01672, 0.01662, 0.01654          & \textbf{0.01628, 0.01642, 0.01634}$^*$ \\
2025                   & 3,751  & 8 & \textbf{0.01655, 0.01637, 0.01665} & 0.01667, 0.01678, 0.01648 \\
2019, 2023, 2025       & 12,853 & 8 & 0.01563, 0.01569, 0.01583          & \textbf{0.01508, 0.01520, 0.01519}$^*$ \\
2019, 2020, 2023       & 12,595 & 8 & 0.01584, 0.01583, 0.01591          & \textbf{0.01510, 0.01517, 0.01513}$^*$ \\
2020, 2022, 2024       & 12,423 & 8 & 0.01568, 0.01570, 0.01558          & \textbf{0.01497, 0.01508, 0.01502}$^*$ \\
2022, 2024, 2025       & 12,681 & 8 & 0.01560, 0.01547, 0.01559          & \textbf{0.01519, 0.01504, 0.01504}$^*$ \\
2019--2020, 2022--2025 & 25,276 & 8 & 0.01534, 0.01517, 0.01524          & \textbf{0.01459, 0.01464, 0.01456}$^*$ \\
2019--2020, 2022--2025 & 1,000  & 8 & \textbf{0.01640, 0.01643, 0.01662}$^*$ & 0.01756, 0.01769, 0.01755 \\
2019--2020, 2022--2025 & 2,000  & 8 & \textbf{0.01598, 0.01611, 0.01601}$^*$ & 0.01682, 0.01651, 0.01637 \\
2019--2020, 2022--2025 & 3,500  & 8 & \textbf{0.01572, 0.01574, 0.01564} & 0.01565, 0.01628, 0.01608 \\
2019--2020, 2022--2025 & 5,000  & 8 & \textbf{0.01562, 0.01553, 0.01544} & 0.01555, 0.01550, 0.01557 \\
2019--2020, 2022--2025 & 8,000  & 8 & 0.01544, 0.01532, 0.01546 & \textbf{0.01498, 0.01500, 0.01519}$^*$ \\
2019--2020, 2022--2025 & 12,500 & 8 & 0.01529, 0.01536, 0.01532          & \textbf{0.01488, 0.01480, 0.01501}$^*$ \\
\addlinespace
2019--2020, 2022--2025 & 25,276 & 2 & -- & 0.01481, 0.01467, 0.01497 \\
2019--2020, 2022--2025 & 25,276 & 4 & -- & 0.01486, 0.01460, 0.01453 \\
2019--2020, 2022--2025 & 25,276 & 6 & -- & 0.01451, \textcolor{green}{0.01446}, 0.01451 \\
\bottomrule
\end{tabular}
\end{table}

\clearpage
\bibliographystyle{plainnat}
\bibliography{bibli_2}

@article{Kopp_et_al_2024,
AUTHOR = {Kopp, J. and Hering, A. and Germann, U. and Martius, O.},
TITLE = {Verification of weather-radar-based hail metrics with crowdsourced observations from Switzerland},
JOURNAL = {Atmospheric Measurement Techniques},
VOLUME = {17},
YEAR = {2024},
NUMBER = {14},
PAGES = {4529--4552},
URL = {https://amt.copernicus.org/articles/17/4529/2024/},
DOI = {10.5194/amt-17-4529-2024}
}

@article{Yousefnia_et_al_2024,
author = {Vahid Yousefnia, Kianusch and Bölle, Tobias and Zöbisch, Isabella and Gerz, Thomas},
title = {A machine-learning approach to thunderstorm forecasting through post-processing of simulation data},
journal = {Quarterly Journal of the Royal Meteorological Society},
volume = {150},
number = {763},
pages = {3495-3510},
doi = {https://doi.org/10.1002/qj.4777},
url = {https://rmets.onlinelibrary.wiley.com/doi/abs/10.1002/qj.4777},
eprint = {https://rmets.onlinelibrary.wiley.com/doi/pdf/10.1002/qj.4777},
year = {2024}
}

@article{Leinonen_et_al_2023,
author = {Leinonen, Jussi and Hamann, Ulrich and Sideris, Ioannis V. and Germann, Urs},
title = {Thunderstorm Nowcasting With Deep Learning: A Multi-Hazard Data Fusion Model},
journal = {Geophysical Research Letters},
volume = {50},
number = {8},
pages = {e2022GL101626},
doi = {https://doi.org/10.1029/2022GL101626},
url = {https://agupubs.onlinelibrary.wiley.com/doi/abs/10.1029/2022GL101626},
eprint = {https://agupubs.onlinelibrary.wiley.com/doi/pdf/10.1029/2022GL101626},
note = {e2022GL101626 2022GL101626},
year = {2023}
}

@article {Leinonen_et_al_2022,
      author = "Jussi Leinonen and Ulrich Hamann and Urs Germann",
      title = "Seamless Lightning Nowcasting with Recurrent-Convolutional Deep Learning",
      journal = "Artificial Intelligence for the Earth Systems",
      year = "2022",
      publisher = "American Meteorological Society",
      address = "Boston MA, USA",
      volume = "1",
      number = "4",
      doi = "10.1175/AIES-D-22-0043.1",
      pages=      "e220043",
      url = "https://journals.ametsoc.org/view/journals/aies/1/4/AIES-D-22-0043.1.xml"
}

@Article{Rombeek_et_al_2024,
AUTHOR = {Rombeek, N. and Leinonen, J. and Hamann, U.},
TITLE = {Exploiting radar polarimetry for nowcasting thunderstorm hazards using deep learning},
JOURNAL = {Natural Hazards and Earth System Sciences},
VOLUME = {24},
YEAR = {2024},
NUMBER = {1},
PAGES = {133--144},
URL = {https://nhess.copernicus.org/articles/24/133/2024/},
DOI = {10.5194/nhess-24-133-2024}
}

@article{Shi_et_al_2015,
  author       = {Xingjian Shi and
                  Zhourong Chen and
                  Hao Wang and
                  Dit{-}Yan Yeung and
                  Wai{-}Kin Wong and
                  Wang{-}chun Woo},
  title        = {Convolutional {LSTM} Network: {A} Machine Learning Approach for Precipitation
                  Nowcasting},
  journal      = {CoRR},
  volume       = {abs/1506.04214},
  year         = {2015},
  url          = {http://arxiv.org/abs/1506.04214},
  eprinttype    = {arXiv},
  eprint       = {1506.04214},
  bibsource    = {dblp computer science bibliography, https://dblp.org}
}

@article{Espeholt_et_al_2022,
  author    = {Lasse Espeholt and Shreya Agrawal and Casper Sønderby and Manoj Kumar and Jonathan Heek and Carla Bromberg and Cenk Gazen and Rob Carver and Marcin Andrychowicz and Jason Hickey and Aaron Bell and Nal Kalchbrenner},
  title     = {Deep learning for twelve hour precipitation forecasts},
  journal   = {Nature Communications},
  year      = {2022},
  volume    = {13},
  number    = {1},
  pages     = {5145},
  doi       = {10.1038/s41467-022-32483-x},
  url       = {https://doi.org/10.1038/s41467-022-32483-x}, 
  issn      = {2041-1723}
}

@Article{Franch_et_al_2020,
AUTHOR = {Franch, Gabriele and Nerini, Daniele and Pendesini, Marta and Coviello, Luca and Jurman, Giuseppe and Furlanello, Cesare},
TITLE = {Precipitation Nowcasting with Orographic Enhanced Stacked Generalization: Improving Deep Learning Predictions on Extreme Events},
JOURNAL = {Atmosphere},
VOLUME = {11},
YEAR = {2020},
NUMBER = {3},
ARTICLE-NUMBER = {267},
URL = {https://www.mdpi.com/2073-4433/11/3/267},
ISSN = {2073-4433},
DOI = {10.3390/atmos11030267}
}

@article{Lam_et_al_2023,
author = {Remi Lam  and Alvaro Sanchez-Gonzalez  and Matthew Willson  and Peter Wirnsberger  and Meire Fortunato  and Ferran Alet  and Suman Ravuri  and Timo Ewalds  and Zach Eaton-Rosen  and Weihua Hu  and Alexander Merose  and Stephan Hoyer  and George Holland  and Oriol Vinyals  and Jacklynn Stott  and Alexander Pritzel  and Shakir Mohamed  and Peter Battaglia },
title = {Learning skillful medium-range global weather forecasting},
journal = {Science},
volume = {382},
number = {6677},
pages = {1416-1421},
year = {2023},
doi = {10.1126/science.adi2336},
URL = {https://www.science.org/doi/abs/10.1126/science.adi2336},
eprint = {https://www.science.org/doi/pdf/10.1126/science.adi2336}}

@book{Markowski_Richardson_2010,
author = {Markowski, Paul and Richardson, Yvette},
address = {Somerset},
edition = {1. Aufl.},
isbn = {0470742135},
language = {eng},
publisher = {Wiley},
series = {Advancing weather and climate science},
title = {Mesoscale Meteorology in Midlatitudes},
volume = {2},
year = {2010},
}

@misc{Meteo_Swiss_OGD,
  author       = {{MeteoSwiss}},
  title        = {Open Data Documentation},
  year         = {2025},
  url          = {https://opendatadocs.meteoswiss.ch/},
  note         = {Accessed: 2025-10-30}
}

@misc{Miralles_et_al_2025,
      title={Observation-guided Interpolation Using Graph Neural Networks for High-Resolution Nowcasting in Switzerland}, 
      author={Ophélia Miralles and Daniele Nerini and Jonas Bhend and Baudouin Raoult and Christoph Spirig},
      year={2025},
      eprint={2509.00017},
      archivePrefix={arXiv},
      primaryClass={physics.ao-ph},
      url={https://arxiv.org/abs/2509.00017}, 
}

@inproceedings{Hering_et_al_2008,
  author       = {Hering, A. M. and Germann, U. and Boscacci, M. and S{\'e}n{\'e}si, S.},
  title        = {Operational nowcasting of thunderstorms in the Alps during MAP D-PHASE},
  booktitle    = {Proceedings of the 5th European Conference on Radar in Meteorology and Hydrology},
  address      = {Helsinki, Finland},
  date         = {2008-06-30/2008-07-04},
  year         = {2008},
}

@inproceedings{Hamann_et_al_2025,
  author    = {Ulrich Hamann and Luca Nisi and Irina Mahlstein and Matteo Buzzi and Michele Cattaneo and N{\'e}stor Tarin Burriel and Przemyslaw Juda and Nathalie Rombeek and George Pacey and Ophélia Miralles and Jussi Leinonen},
  title     = {Nowcasting of Thunderstorm Hazards with Deep Learning: Performance Report of the First Convective Season in Operations},
  booktitle = {12th European Conference on Severe Storms (ECSS2025)},
  year      = {2025},
  address   = {Utrecht, The Netherlands},
  month     = {Nov},
  day       = {17--21},
  note      = {ECSS2025-126},
  doi       = {10.5194/ecss2025-126}
}

@misc{GVL_2022,
  author       = {{GVL}},
  title        = {Hagelereignis überschattet das Geschäftsjahr 2021},
  howpublished = {\emph{Media release}},
  year         = {2022},
  month        = {May},
  day          = {5},
  note         = {Accessed from \url{https://www.gvl.ch/downloads/Geschaeftsjahr_2021_-_Medienmitteilung.pdf}}
}

@misc{lang_et_al_2024_aif,
      title={AIFS -- ECMWF's data-driven forecasting system}, 
      author={Simon Lang and Mihai Alexe and Matthew Chantry and Jesper Dramsch and Florian Pinault and Baudouin Raoult and Mariana C. A. Clare and Christian Lessig and Michael Maier-Gerber and Linus Magnusson and Zied Ben Bouallègue and Ana Prieto Nemesio and Peter D. Dueben and Andrew Brown and Florian Pappenberger and Florence Rabier},
      year={2024},
      eprint={2406.01465},
      archivePrefix={arXiv},
      primaryClass={physics.ao-ph},
      url={https://arxiv.org/abs/2406.01465}, 
}

@article{kopp2023summer,
	title        = {The summer 2021 {S}witzerland hailstorms: weather situation, major impacts and unique ­observational data},
	author       = {Kopp, Jérôme and Schröer, Katharina and Schwierz, Cornelia and Hering, Alessandro and Germann, Urs and Martius, Olivia},
	year         = 2023,
	journal      = {Weather},
	volume       = 78,
	number       = 7,
	pages        = {184--191},
	doi          = {https://doi.org/10.1002/wea.4306},
	url          = {https://rmets.onlinelibrary.wiley.com/doi/abs/10.1002/wea.4306}
}

@article{nisi2018alpine,
	title        = {A 15-year hail streak climatology for the {A}lpine region},
	author       = {Nisi, Luca and Hering, Alessandro and Germann, Urs and Martius, Olivia},
	year         = 2018,
	journal      = {Quarterly Journal of the Royal Meteorological Society},
	volume       = 144,
	number       = 714,
	pages        = {1429--1449},
	doi          = {https://doi.org/10.1002/qj.3286},
	url          = {https://rmets.onlinelibrary.wiley.com/doi/abs/10.1002/qj.3286}
}

@article{Wapler_and_James_2015,
title = {Thunderstorm occurrence and characteristics in Central Europe under different synoptic conditions},
journal = {Atmospheric Research},
volume = {158-159},
pages = {231-244},
year = {2015},
issn = {0169-8095},
doi = {https://doi.org/10.1016/j.atmosres.2014.07.011},
url = {https://www.sciencedirect.com/science/article/pii/S0169809514002749},
author = {Kathrin Wapler and Paul James}
}

@InProceedings{Ronneberger_et_al_2015,
author="Ronneberger, Olaf
and Fischer, Philipp
and Brox, Thomas",
editor="Navab, Nassir
and Hornegger, Joachim
and Wells, William M.
and Frangi, Alejandro F.",
title="U-Net: Convolutional Networks for Biomedical Image Segmentation",
booktitle="Medical Image Computing and Computer-Assisted Intervention -- MICCAI 2015",
year="2015",
publisher="Springer International Publishing",
address="Cham",
pages="234--241",
isbn="978-3-319-24574-4"
}

@misc{Ballas_et_al_2016,
      title={Delving Deeper into Convolutional Networks for Learning Video Representations}, 
      author={Nicolas Ballas and Li Yao and Chris Pal and Aaron Courville},
      year={2016},
      eprint={1511.06432},
      archivePrefix={arXiv},
      primaryClass={cs.CV},
      url={https://arxiv.org/abs/1511.06432}, 
}

@misc{Andrychowicz_et_al_2023,
      title={Deep Learning for Day Forecasts from Sparse Observations}, 
      author={Marcin Andrychowicz and Lasse Espeholt and Di Li and Samier Merchant and Alexander Merose and Fred Zyda and Shreya Agrawal and Nal Kalchbrenner},
      year={2023},
      eprint={2306.06079},
      archivePrefix={arXiv},
      primaryClass={physics.ao-ph},
      url={https://arxiv.org/abs/2306.06079}, 
}

@Article{Wilhelm_et_al_2024,
AUTHOR = {Wilhelm, L. and Schwierz, C. and Schr\"oer, K. and Taszarek, M. and Martius, O.},
TITLE = {Reconstructing hail days in Switzerland with statistical models (1959--2022)},
JOURNAL = {Natural Hazards and Earth System Sciences},
VOLUME = {24},
YEAR = {2024},
NUMBER = {11},
PAGES = {3869--3894},
URL = {https://nhess.copernicus.org/articles/24/3869/2024/},
DOI = {10.5194/nhess-24-3869-2024}
}

@inproceedings{Foote_et_al_2005,
  author    = {Foote, G. Brant and Krauss, Terrence W. and Makitov, Viktor},
  title     = {Hail Metrics Using Conventional Radar},
  booktitle = {85th AMS Annual Meeting},
  year       = {2005},
  address    = {San Diego, CA, USA},
  publisher  = {American Meteorological Society},
  pages      = {2791--2796},
  url        = {https://scholars.duke.edu/publication/759157},
  note       = {8--14 January 2005}
}

@article{Waldvogel_et_al_1979,
  author  = {Waldvogel, A. and Federer, B. and Grimm, P.},
  title   = {Criteria for the Detection of Hail Cells},
  journal = {Journal of Applied Meteorology},
  year    = {1979},
  volume  = {18},
  number  = {12},
  pages   = {1521--1525},
  doi     = {10.1175/1520-0450(1979)018<1521:CFTDOH>2.0.CO;2}
}

@article{LeCun_et_al_1998,
  title={Gradient-Based Learning Applied to Document Recognition},
  author={LeCun, Yann and Bottou, L{\'e}on and Bengio, Yoshua and Haffner, Patrick},
  journal={Proceedings of the IEEE},
  volume={86},
  number={11},
  pages={2278--2324},
  year={1998},
  publisher={IEEE}
}

@Article{Franch_et_al_2025,
AUTHOR = {Franch, G. and Tomasi, E. and Wanjari, R. and Poli, V. and Cardinali, C. and Alberoni, P. P. and Cristoforetti, M.},
TITLE = {GPTCast: a weather language model for precipitation nowcasting},
JOURNAL = {Geoscientific Model Development},
VOLUME = {18},
YEAR = {2025},
NUMBER = {16},
PAGES = {5351--5371},
URL = {https://gmd.copernicus.org/articles/18/5351/2025/},
DOI = {10.5194/gmd-18-5351-2025}
}

@article{Kunz_et_al_2018,
author = {Kunz, Michael and Blahak, Ulrich and Handwerker, Jan and Schmidberger, Manuel and Punge, Heinz Jürgen and Mohr, Susanna and Fluck, Elody and Bedka, Kris M.},
title = {The severe hailstorm in southwest Germany on 28 July 2013: characteristics, impacts and meteorological conditions},
journal = {Quarterly Journal of the Royal Meteorological Society},
volume = {144},
number = {710},
pages = {231-250},
doi = {10.1002/qj.3197},
url = {https://rmets.onlinelibrary.wiley.com/doi/abs/10.1002/qj.3197},
eprint = {https://rmets.onlinelibrary.wiley.com/doi/pdf/10.1002/qj.3197},
year = {2018}
}

@article {McGovern_et_al_2023,
      author = "Amy McGovern and Randy J. Chase and Montgomery Flora and David J. Gagne and Ryan Lagerquist and Corey K. Potvin and Nathan Snook and Eric Loken",
      title = "A Review of Machine Learning for Convective Weather",
      journal = "Artificial Intelligence for the Earth Systems",
      year = "2023",
      publisher = "American Meteorological Society",
      address = "Boston MA, USA",
      volume = "2",
      number = "3",
      doi = "10.1175/AIES-D-22-0077.1",
      pages=      "e220077",
      url = "https://journals.ametsoc.org/view/journals/aies/2/3/AIES-D-22-0077.1.xml"
}

@inproceedings{boloni_et_al_2025,
  author    = {Bölöni, G. and James, P. M. and Debertshäuser, M. and Theis, S.},
  title     = {Towards a machine-learning enhanced nowcasting tool for storm severity analysis and prediction},
  booktitle = {12th European Conference on Severe Storms},
  address   = {Utrecht, The Netherlands},
  year      = {2025},
  month     = nov,
  pages     = {17--21},
  number    = {ECSS2025-158},
  doi       = {10.5194/ecss2025-158}
}

@inproceedings{montmerle_et_al_2025,
  author    = {Montmerle, T. and Tzanos, R. and Pottez, E. and Arnould, G. and Jaubert, D. and Moisselin, J.-M.},
  title     = {Recent work at Météo-France's Nowcasting department on warnings of exceptional rainfall and thunderstorm objects},
  booktitle = {12th European Conference on Severe Storms},
  address   = {Utrecht, The Netherlands},
  year      = {2025},
  month     = nov,
  pages     = {17--21},
  number    = {ECSS2025-225},
  doi       = {10.5194/ecss2025-225}
}

@misc{swisstopo_2024,
  author       = {{Federal Office of Topography swisstopo}},
  title        = {The Swiss coordinates system},
  year         = {2024},
  howpublished = {\url{https://www.swisstopo.admin.ch/en/the-swiss-coordinates-system}},
  note         = {Published 8 January 2024},
}

@article {Lagerquist_et_al_2020,
      author = "Ryan Lagerquist and Amy McGovern and Cameron R. Homeyer and David John Gagne II and Travis Smith",
      title = "Deep Learning on Three-Dimensional Multiscale Data for Next-Hour Tornado Prediction",
      journal = "Monthly Weather Review",
      year = "2020",
      publisher = "American Meteorological Society",
      address = "Boston MA, USA",
      volume = "148",
      number = "7",
      doi = "10.1175/MWR-D-19-0372.1",
      pages=      "2837 - 2861",
      url = "https://journals.ametsoc.org/view/journals/mwre/148/7/mwrD190372.xml"
}

@Article{Tran_and_Song_2019,
AUTHOR = {Tran, Quang-Khai and Song, Sa-kwang},
TITLE = {Multi-Channel Weather Radar Echo Extrapolation with Convolutional Recurrent Neural Networks},
JOURNAL = {Remote Sensing},
VOLUME = {11},
YEAR = {2019},
NUMBER = {19},
ARTICLE-NUMBER = {2303},
URL = {https://www.mdpi.com/2072-4292/11/19/2303},
ISSN = {2072-4292},
DOI = {10.3390/rs11192303}
}

\end{document}